\documentclass[conference]{IEEEtran}

\ifCLASSINFOpdf
\else
\fi

\usepackage{amsmath}
\usepackage{amssymb}
\usepackage{amsfonts}
\usepackage{mathrsfs}
\usepackage{color}

\usepackage{graphicx}
\usepackage{subfigure}
\usepackage{float}

\begin{document}
\title{Stopping Condition for Greedy Block Sparse Signal Recovery}

\author{
Yu Luo, Ronggui Xie, Huarui Yin, and Weidong Wang\\
\small Department of Electronics Engineering and Information Science,
University of Science and Technology of China\\
Email: \{luoyu06, xrghit03\}@mail.ustc.edu.cn, \{yhr, wdwang\}@ustc.edu.cn}

\maketitle

\begin{abstract}
For greedy block sparse recovery where the sparsity level is unknown,
we derive a stopping condition to stop the iteration process.
Focused on the block orthogonal matching pursuit (BOMP) algorithm,
we model the energy of residual signals at each iteration from a probabilistic
perspective. At the iteration when the last supporting block is detected, the
resulting energy of residual signals is supposed to suffer an obvious decrease.
Based on this, we stop the iteration process when the energy of residual signals
is below a given threshold. Compared with other approaches, our derived
condition works well for the BOMP recovery. What is more, we promote
our approach to the interference cancellation based BOMP (ICBOMP) recovery in
paper \cite{xie2015many}. Simulation results show that our derived condition
can save many unnecessary iterations and at the same time guarantees a
favorable recovery accuracy, both for the BOMP and ICBOMP recoveries.
\end{abstract}


\IEEEpeerreviewmaketitle

\section{Introduction}
In the last few years, compressed sensing (CS) \cite{candes2008introduction}
has drawn increased interest in many areas such as signal processing and multi-user communications \cite{zhu2011exploiting,bockelmann2013compressive,zhang2011neighbor}.
The CS theory claims that when the signals of interest are sparse with many elements being zero,
even sampling the signals using a rate less than the Nyquist rate, it
can be recovered from the down-sampled measurements almost without
losing the information. The early work on CS assumes that each of the nonzero
signals is just randomly located among all possible positions of a vector,
i.e., random-sparsity case. However, as stated in papers such as \cite{eldar2010block},
the nonzero signals are usually clustered, exhibiting the structure of block-sparsity.
The block-sparsity indicates that, when partitioning the sequential signals
into blocks, only some blocks contain nonzero components and all other blocks are zero.

Suppose ${\bf{s}}$ is an $Nd \times 1$ signal vector given as ${\bf{s}} =
{\left[ {{\bf{s}}_1^T \cdots ,{\bf{s}}_i^T \cdots ,{\bf{s}}_N^T} \right]^T}$
where superscript $T$ stands for the transpose and ${{\bf{s}}_i}$  is a $d \times 1$ sub-vector,
$1 \le i \le N$. Suppose only $N_a$ out of $N$ sub-vectors are nonzero, usually with $N_a \ll N$.
The sparsity level is thus $N_a$. When $d = 1$, ${\bf{s}}$ exhibits the property of
random-sparsity, and when $d > 1$, ${\bf{s}}$  exhibits the structure of block-sparsity.
The CS measures ${\bf{s}}$  using an $M \times Nd$ measurement matrix
${\bf{B}} = \left[ {{{\bf{B}}_1},{{\bf{B}}_2}, \cdots ,{{\bf{B}}_N}} \right]$,
given as ${\bf{y}} = {\bf{Bs}}$, with $M < Nd$, where $M$ stands for the measurement number. If the measurement is performed in a
noisy environment, it has ${\bf{y}} = {\bf{Bs}} + {\bf{z}}$ where ${\bf{z}}$ represents the noise vector.

For sparse signal recovery, many algorithms such as
\cite{tropp2007signal,needell2009cosamp}
are proposed. Among all the algorithms, greedy algorithms \cite{tropp2007signal,needell2009cosamp}
are important since they are simple for practical use. All the greedy recovery algorithms of random-sparsity
can be transplanted to the block-sparsity case. For example, the block OMP (BOMP) is developed
from the OMP algorithm for the block-sparse recovery \cite{eldar2010block}.
Existing results demonstrate that, compared with the random-sparsity, exploiting the block
structure provides better signal reconstruction performance.

Sparsity level $N_a$ is an important parameter
for the sparse recovery, especially for
the greedy recovery. Many works, such as
\cite{tropp2007signal,needell2009cosamp} assume
that the $N_a$ is a priori known to control
the iteration number. Unfortunately in reality, $N_a$ is usually
unknown at the signal recovery side and its
estimation is therefore necessary. It should be
noted that, if the estimated sparsity level
is smaller than $N_a$, some nonzero signals will
certainly be missed to detect; if the estimated
value is larger than $N_a$, unnecessary iterations
will harm the recovery performance \cite{do2008sparsity},
including degradation in accuracy and
increase in complexity. To address this problem,
work in \cite{qiu2012sparse} proposes automatic
double overrelaxation (ADORE) thresholding method
to estimate the sparsity level and reconstruct
the signal simultaneously. Other works such as
\cite{do2008sparsity,cai2011orthogonal}
also adopt some stop criterions to stop the iterations
process of the greedy recovery. However, all the above works
are for the random-sparse recovery.

In this paper, we focus on the BOMP recovery of the block-sparsity situation
where the sparsity level is unknown.
Rather than giving the stopping condition by experience, or setting a maximum
iteration number as in \cite{ben2011near}, we theoretically derive the stopping condition.
We model the energy of the residual signal vector from a probabilistic perspective and
we use its distribution to derive a threshold to stop the greedy process. When the energy of residual signal is smaller than that threshold,
all the supporting blocks are supposed to have been detected and the BOMP algorithm
will finish its iteration process. This approach works well for the BOMP,
as demonstrated by the simulation. This gives us the confidence to promote the method.
Specially, we use the same method to derive the stopping condition for the iterations
of interference cancellation based BOMP (ICBOMP) algorithm in \cite{xie2015many}.
The ICBOMP is developed from the BOMP algorithm for the small packet recovery.

The rest of the paper is organized as follows. In Section II, we derive
the iteration stopping condition for the BOMP recovery. In Section III,
we transplant the method to the ICBOMP recovery in \cite{xie2015many}.
In section IV, some related works are cited. Finally, numerical
studies are shown in Section V, followed by the conclusion in Section VI.

Notation: Vectors and matrices are denoted by boldface lowercase and uppercase letters,
respectively. The identity matrix of size $n \times n$ is denoted as ${{\bf{I}}_n}$.
For a subset $I \subset \left[ N \right]: = \left\{ {1,2, \cdots ,N} \right\}$ and a
matrix  ${\bf{B}}: = \left[ {{{\bf{B}}_1},{{\bf{B}}_2}, \cdots ,{{\bf{B}}_N}} \right]$
consisting of $N$ sub-matrices (blocks) of equal size,
${{\bf{B}}_I}$ stands for a sub-matrix of  ${\bf{B}}$ whose block indices are in set $I$;
for a vector ${\bf{s}}: = {[{\bf{s}}_1^T,{\bf{s}}_2^T, \cdots ,{\bf{s}}_N^T]^T}$,  ${{\bf{s}}_I}$
is similarly defined. Value $\left| I \right|$  stands for the cardinality of set $I$.
Given two sets ${I_1}$  and ${I_2}$, ${I_1}\backslash {I_2} = {I_1} \cap I_2^c$. $ \otimes $ stands for the Kronecker product.

\section{Stopping Condition for the BOMP Recovery}

In this part, we give a more detailed description for the block-sparsity recovery problem,
and we take the BOMP algorithm as an example to derive the stopping condition from the probabilistic perspective.

\subsection{Block-sparsity Recovery Problem}
As mentioned earlier, the measurement of block-sparse signal vector $\bf{s}$ in a noisy environment is given as
$\bf{y} = {\bf{Bs}} + {\bf{z}}$. In this paper, all the parameters are assumed in complex field. Besides,
for the later derivation convenience, we assume that: 1) matrix ${\bf{B}}$  is randomly generated,
and all its entries are i.i.d. Gaussian variables with a mean zero and a variance ${1 \over M}$; 2) nonzero signals in $\bf{s}$
are i.i.d. variables of zero mean and unit variance;
3) noise ${\bf{z}} \sim \mathcal {CN} \left( {{\bf{0}},{\sigma ^2}{{\bf{I}}_{M}}} \right)$.
Gaussian approach makes the ${\bf{B}}$ almost surely satisfy the
restricted isometry property (RIP) which is necessary for the sparse recovery \cite{baraniuk2010model}.

Let $I$  be the set containing the unknown indices of $N_a$ nonzero blocks,
with cardinal number $\left| I \right| = {N_a}$. Then the measurement can be rewritten as
\begin{equation}
{\bf{y}} = \sum_{i \in I} {{{\bf{B}}_i}{{\bf{s}}_i} + } {\bf{z}} = {{\bf{B}}_I}{{\bf{s}}_I} + {\bf{z}}.
\end{equation}

\subsection{BOMP Recovery}
Also for the later derivation convenience, the iteration process of the BOMP is summarized here.
At the $k$-th iteration, $k \in \left\{ {1, 2, \cdots } \right\}$, let ${{\bf{r}}_k}$
denote the residual signal and ${\Lambda _k}$ denote the set containing the indices of already detected blocks,
their initializations are respectively given as ${{\bf{r}}_0} = {\bf{y}}$ and ${\Lambda _0} = \emptyset $.
Then, the BOMP iteration is performed as follows:

1) For  $j \in \left\{ {\left[ N \right]\backslash {\Lambda _{k - 1}}} \right\}$, select the block
that has the maximum correlation with the residual signals:
$${j_k} = \mathop {\arg \max }\limits_j \left\| {{\bf{B}}_j^H{{\bf{r}}_{k - 1}}} \right\|_2^2$$

2) update the index set:
$${\Lambda _k} = {\Lambda _{k - 1}} \cup \left\{ {{j_k}} \right\}$$

3) update the signal by the least-square (LS) algorithm:
$${{\bf{\bar s}}_{{\Lambda _k}}} = \mathop {\arg \min }\limits_{{{\bf{s}}_0}} {\left\| {{\bf{y}} - {{\bf{B}}_{{\Lambda _k}}}{{\bf{s}}_0}} \right\|_2}$$

4) update the residual signals:
$${{\bf{r}}_k} = {\bf{y}} - {{\bf{B}}_{{\Lambda _k}}}{{\bf{\bar s}}_{{\Lambda _k}}}$$

The above BOMP iterations are terminated when certain condition is satisfied, either
it reaches to the maximum allowed iteration number as in \cite{ben2011near,xie2015many}, or the
energy of the residual signals is below an empirical value as in \cite{mallat1993matching}.
The later approach is based on the common fact that
the energy of the residual signals will usually suffer and obvious decrease
when the last supporting block is selected. Different from these two kinds of approaches,
in the following, by viewing the energy of the residual signals to be a random variable,
we theoretically derive the iteration stopping condition from a probabilistic perspective.

\subsection{Energy Evaluation of Residual Signal}
At the $k$-$th$ BOMP iteration, it has  $|{\Lambda _k}| = k$. The signal update is given as
\begin{equation}
{{\bf{\bar s}}_{{\Lambda _k}}} = {({\bf{B}}_{{\Lambda _k}}^H{{\bf{B}}_{{\Lambda _k}}})^{ - 1}}{\bf{B}}_{{\Lambda _k}}^H{\bf{y}}.
\end{equation}
The energy of residual signal is a random variable and is defined as ${E_k} = \left\| {{{\bf{r}}_k}} \right\|_2^2
= {\left( {{\bf{y}} - {{\bf{B}}_{{\Lambda _k}}}{{{\bf{\bar s}}}_{{\Lambda _k}}}} \right)^H}
\left( {{\bf{y}} - {{\bf{B}}_{{\Lambda _k}}}{{{\bf{\bar s}}}_{{\Lambda _k}}}} \right)$.
Assume that there are $n_a$ $\left( {0 \le n_a \le {N_a}} \right)$ supporting blocks
remained to detect, i.e., $\left| {I\backslash {\Lambda _k}} \right| = n_a$, then the ${E_k}$
has a mean value given as follows
\begin{equation}  \label{mean}
 \begin{aligned}
{\mu _k} = & {\bf{E}}\left[ {{\bf{s}}_{I\backslash {{\rm{\Lambda }}_{\rm{k}}}}^H{\bf{B}}_{I\backslash {{\rm{\Lambda }}_{\rm{k}}}}^H{{\bf{B}}_{I\backslash {{\rm{\Lambda }}_{\rm{k}}}}}{{\bf{s}}_{I\backslash {{\rm{\Lambda }}_{\rm{k}}}}}} \right]\\
           &- {\bf{E}}\left[ {{\bf{s}}_{I\backslash {{\rm{\Lambda }}_{\rm{k}}}}^H{\bf{B}}_{I\backslash {{\rm{\Lambda }}_{\rm{k}}}}^H{{\bf{B}}_{{{\rm{\Lambda }}_{\rm{k}}}}}{{({\bf{B}}_{{{\rm{\Lambda }}_{\rm{k}}}}^H{{\bf{B}}_{{{\rm{\Lambda }}_{\rm{k}}}}})}^{ - 1}}{\bf{B}}_{{{\rm{\Lambda }}_{\rm{k}}}}^H{{\bf{B}}_{I\backslash {{\rm{\Lambda }}_{\rm{k}}}}}{{\bf{s}}_{I\backslash {{\rm{\Lambda }}_{\rm{k}}}}}} \right]\\
           &+ {\bf{E}}\left[ {{{\bf{z}}^H}({{\bf{I}}_M} - {{\bf{B}}_{{{\rm{\Lambda }}_{\rm{k}}}}}{{({\bf{B}}_{{{\rm{\Lambda }}_{\rm{k}}}}^H{{\bf{B}}_{{{\rm{\Lambda }}_{\rm{k}}}}})}^{ - 1}}{\bf{B}}_{{{\rm{\Lambda }}_{\rm{k}}}}^H){\bf{z}}} \right] \\
           =& {n_a}d - {n_a}{{k{d^2}} \over {{M}}}{\rm{ + }}\left( {{M} - kd} \right){\sigma ^2}\\
           =&\left( {M - kd} \right)\left( {{\sigma ^2} + {{{n_a}d} \over M}} \right)
\end{aligned}
\end{equation}
where the property of the mathematical trace operation is used.
And it should be noted that a more exact mean value should consider the order statistics of signal blocks,
but the expressions will be complicated. For deriving a usable mean value, the above derivations omit the order statistics.

Since each component of  ${{{\bf{r}}_k}}$ is a superposition of many independent variables, it can be
approximated as a Gaussian variable. We further assume that components of ${{{\bf{r}}_k}}$ are i.i.d.
Gaussian variables and each of them has a mean of zero and a variance of  ${{\tilde \sigma }^2}$, with
${{\tilde \sigma }^2} = {{{\mu _k}} \over M} = {{M - kd} \over M}\left( {{\sigma ^2} + {{{n_a}d} \over M}} \right)$.
Then ${E_k}$ follows a chi-squared distribution with $2M$ degrees of freedom, and its
variance is given as
\begin{equation}  \label{variance}
\begin{aligned}
\sigma _k^2 = & M(M+1)\left({{\tilde \sigma }^2}\right)^2-\mu _k^2\\
           =&{{{{\left( {M - kd} \right)}^2}} \over M}{\left( {{\sigma ^2} + {{{n_a}d} \over M}} \right)^2}
\end{aligned}
\end{equation}

Usually, $M$ is large. In this case, it's reasonable to treat ${E_k}$ as a Gaussian variable,
satisfying ${E_k} \sim \mathcal{N}\left( {{\mu _k},\sigma _k^2} \right)$.

\subsection{Stopping Condition}   \label{II-D}
As above stated, when the last supporting block is selected at the $k$-th iteration of BOMP algorithm,
a sharp decrease will happen to the energy of the residual signals. This gives us
the idea to derive a threshold, to stop the BOMP iterations.
That is if $E_k$ is smaller than the set threshold, the last supporting
block is supposed to have been selected and then the iterations can be terminated.

The $E_k$ is a random variable, and its distribution is decided by the following two cases:

\textbf{C1}: $\left| {I\backslash {\Lambda _k}} \right| = n_a \ge 1$.

\textbf{C0}: $\left| {I\backslash {\Lambda _k}} \right| = n_a = 0$.

The mean and variance of the ${E_k}$ are respectively given as
(\ref{mean}) and (\ref{variance}), for both of the above two cases.
When performing energy detection by a threshold $\eta_{k,1}$,
a missed detection probability, say $p_m$, will happen by
deciding the \textbf{C1} to be
the \textbf{C0}. Applying Gaussian variable to approximate $E_k$, it has
that
\begin{equation}  \label{missp}
  P({E_k} \le \eta_{k,1}) = \Phi \left( {{{{\eta_{k,1}} - {\mu _k}} \over {{\sigma _k}}}} \right) = p_m
\end{equation}
where $\Phi \left( x \right) = {1 \over {\sqrt {2\pi } }}\int_{ - \infty }^x {\exp \left( { - {{{t^2}} \over 2}} \right)} dt$.
By substituting (\ref{mean}) and (\ref{variance}) into (\ref{missp}), it gives that
\begin{equation}  \label{threshold1}
  \eta_{k,1} = \left( {M - kd} \right)\left( {{\sigma ^2} + {{n_a}d \over M}} \right)\left( {1 + {{{\Phi ^{ - 1}}\left( {{p_{m}}} \right)} \over {\sqrt M }}} \right)
\end{equation}
where $\Phi^{-1}(p_m)$ is the inverse function of $\Phi(x)$. $\eta_{k,1}$ can be regarded as the maximum threshold for a maximum allowed missed detection probability $p_m$.

On the other hand, if a maximum false detection probability, say $p_f$, is allowed
for deciding the \textbf{C0} to be the \textbf{C1},
it has that
\begin{equation} \label{falsep}
  P({E_k} \ge \eta_{k,0}) = 1- \Phi \left( {{{{\eta_{k,0}} - {\mu _k}} \over {{\sigma _k}}}} \right) = p_f
\end{equation}
which gives that
\begin{equation}   \label{threshold2}
  \eta_{k,0} = \left( {M - kd} \right)\left( {1 - {{{\Phi ^{ - 1}}\left( {{p_{f}}} \right)} \over {\sqrt M }}} \right){\sigma ^2}
\end{equation}

Undoubtedly, if the set threshold, say $\eta_k$, is required to take both the missed
detection probability and false detection probability into account, a
tradeoff should be made between the two probabilities. Note that,
if the false detection happens under the \textbf{C0},
the iteration continues and
some non-supporting blocks will be selected for signal update. This will degrade
the recovery accuracy and at the same time increase the recovery complexity; However, when missed detection happens to the \textbf{C1},
some supporting blocks will be identified to be non-supporting, which will severely
have an adverse impact on the sparse recovery performance.
Therefore, a more accuracy performance cares more about the
missed detection probability. Suppose $p_m$ and $p_f$ are
respectively the allowed missed and false alarm probabilities,
then the reasonable $\eta_k$ is given as follows
\begin{equation}
  \eta_k  = \min \left( {{\eta_{k,1}},{\eta_{k,0}}} \right)
\end{equation}

{\slshape {Remark 1}}: Since iteration is processed at the recovery side,
parameter $k$ can be exactly known at the recovery side. Therefore, threshold
$\eta_k$ will be adjusted with iteration $k$.

{\slshape {Remark 2}}: In practice, we set ${n_a}=1$ to derive $\eta_{k,1}$, because:
1) $n_a$ is unknown at the recovery side, which can not be directly used; 2)
for the threshold derived from ${n_a} = 1$,
conditional probability $P({E_k} \le \eta_k| n_a \ge 2) $ is smaller than conditional probability
$ P({E_k} \le \eta_k| n_a = 1)$, this means the derived threshold
$\eta_k$ is also applicable for the $k$-th iteration when two or more
supporting blocks are remained to detect.

\section{Stopping Condition for the ICBOMP Recovery}
In the communication scenario of \cite{xie2015many}, an uplink system of
$N$ mobile users and a base station (BS) with $M_{\text{ant}}$ antennas is considered.
By exploiting the sparse block transmission that only $N_a$ out of the total $N$
users are actively and simultaneously transmitting data, the work also establishes
the block-sparsity model as follows
\begin{equation}  \label{spmodel}
  {\bf{y}} = \sqrt {{\rho _0}} \sum\limits_{n = 1}^N {{{\bf{B}}_n}{{\bf{s}}_n} + {\bf{z}}}  = \sqrt {{\rho _0}} {\bf{Bs}} + {\bf{z}}
\end{equation}
where ${\rho _0}$ is the signal to noise ratio (SNR).
As a block of ${\bf{B}} \in {\mathbb{C}}^{M_{\text{ant}} T \times Nd}$,
${\bf{B}}_n = {\bf{P}}_n \otimes{\bf{h}}_n \in {\mathbb{C}}^{M_{\text{ant}} T \times d}$
where ${\bf{P}}_n \in {\mathbb{C}^{T \times d}}$ is a kind of
precoding matrix and ${\bf{h}}_n \in {\mathbb{C}^{M_{\text{ant}} \times 1}}$
is the channel gain from the $n$-th user to the BS, $1\le n \le N$.
${\bf{s}}$ is the block-sparse signal to be recovered, with length $d$
for each block ${\bf{s}}_n$. ${\bf{z}}$ is the complex Gaussian noise vector.

To improve the recovery performance, the authors in \cite{xie2015many}
propose the interference cancellation based BOMP (ICBOMP) algorithm, which improves
from the BOMP algorithm by taking advantage of the error correction and detection
code in the communication, to perform the recovery of $\bf{s}$. The ICBOMP
behaves the same as the BOMP in block detection, signal update and residual
update. Their main difference is that for the ICBOMP, some blocks of signals may have been
correctly recovered before finishing all the iterations and need no further update.
However, in \cite{xie2015many} the problem of when to stop the
ICBOMP iterations is not specially studied, the authors only set
a maximum iteration number. In this part, we derive the stopping condition for the ICBOMP
algorithm. For detailed process of the
ICBOMP algorithm, please refer to \cite{xie2015many}.

As the performance analysis in \cite{xie2015many}, entries of ${\bf{P}}_n$,
${\bf{h}}_n$ and ${\bf{z}}$ are all assumed to be i.i.d. complex Gaussian variables, respectively
in $ \mathcal{CN}(0, \frac{1}{T})$, $ \mathcal{CN}(0, 1)$ and $ \mathcal{CN}(0, 1)$.
Nonzero entries of $\bf{s}$ are i.i.d. Quadrature Phase Shift Keying (QPSK) symbols,
each of which has unit energy.
Besides, it should be noted that, by the ICBOMP algorithm, it has $1\le |{\Lambda _k}| = l \le k$.
Suppose $n_a$ active users are remained to detect when the $k$-th iteration
is finished, then similar to the derivations of (\ref{mean}) and (\ref{variance}), the
mean and variance of the residual energy of ICBOMP are respectively given as
\begin{equation}
\begin{aligned}
{\mu _k} &= \left( {M_{\text{ant}}T - ld} \right)\left( {1 + {{{n_a}{\rho _0}d} \over T}} \right)\\
\sigma _k^2 &= {{{{(M_{\text{ant}}T - ld)}^2}} \over {M_{\text{ant}}T}}{\left( {1 + {{{n_a}{\rho _0}d} \over T}} \right)^2}
\end{aligned}
\end{equation}
and the final energy threshold is given by $\eta_k  = \min \left( {{\eta_{k,1}},{\eta_{k,0}}} \right)$,
where the $\eta_{k,1}$ and $\eta_{k,0}$ are respectively given by
\begin{equation}
\begin{aligned}
&\eta_{k,1} = \left( {M_{\text{ant}}T - ld} \right)\left( {1 + {{{n_a}{\rho _0}d} \over T}} \right)\left( {1 + {{{\Phi ^{ - 1}}\left( {{p_m}} \right)} \over {\sqrt {M_{\text{ant}}T} }}} \right)\\
&\eta_{k,0} = \left( {M_{\text{ant}}T - ld} \right)\left( {1 - {{{\Phi ^{ - 1}}\left( {{p_f}} \right)} \over {\sqrt {M_{\text{ant}}T} }}} \right)\\
\end{aligned}
\end{equation}
for certain allowed missed alarm probability $p_m$ and false alarm probability $p_f$.
As the previous Section \ref{II-D}, $n_a =1$ is used to derive $\eta_{k,1}$.

\section{Related works}
In sparse signal recovery literature, many earlier works have
considered the stopping condition for greedy algorithms.
As a conclusion, three common stopping conditions are
\begin{align}
\text{Condition 1} &: {{{\left\| {{{\bf{\bar s}}_{k + 1}} - {{\bf{\bar s}}_k}} \right\|}_2} \over {{{\left\| {{{\bf{\bar s}}_k}} \right\|}_2}}} < {\epsilon _1}  \label{other1} \\
\text{Condition 2} &: \left\| {{{\bf{r}}_k}} \right\|_2^2 < {\epsilon _2} \label{other2} \\
\text{Condition 3} &: \text{setting a maximum iteration number}.
\end{align}

Condition 1 indicates that the algorithm will stop when the relative change of
the reconstructed signals between two consecutive iterations is smaller than a certain value.
This kind of approach is mentioned in \cite{do2008sparsity}, but no specific $\epsilon_1$ is given
in the paper. In \cite{wang2010sparse}, empirical values like $10^{ - 6}$ is set for $\epsilon_1$.
Condition 2 shows that the algorithm will stop when the energy of the
residual signals is smaller than a certain threshold. In \cite{do2008sparsity},
threshold $\epsilon _2$ is set to be the energy of noise vector.
And in \cite{cai2011orthogonal}, such a stopping condition is also theoretically discussed.
Other works like \cite{ben2011near} sets a maximum iteration number, and
\cite{tropp2007signal} assumes that $N_a$ is known and iteration number is exactly set $N_a$.
However, such kinds of approaches are not feasible for practical use, especially when
$N_a$ cannot be a priori known.
It should also be noted that all the above works are for random sparsity case.

In our later numerical studies for the BOMP recovery, Condition 1 and Condition 2
will be simulated for the block sparsity case for comparison. And for the ICBOMP recovery,
Condition 3 will be simulated for comparison.

\begin{figure}
\centering
\subfigure[Iteration number]{
\label{fig:subfig:a}
\includegraphics[width=2in,height=1.8in]{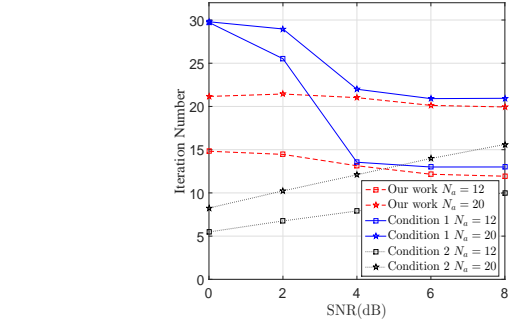}}
\subfigure[NMSE]{
\label{fig:subfig:b}
\includegraphics[width=2in,height=1.8in]{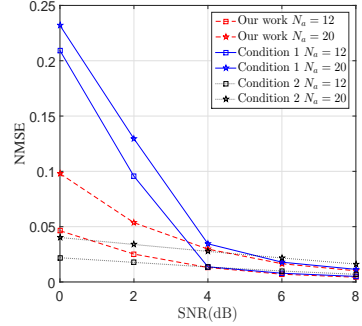}}
\subfigure[Successful detection probability for supporting blocks]{
\label{fig:subfig:c}
\includegraphics[width=2in,height=1.8in]{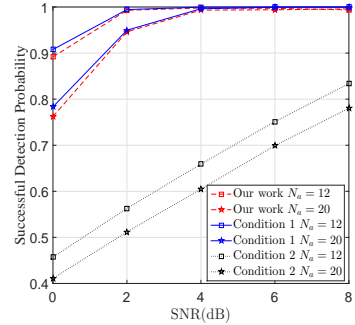}}
\caption{Performance for BOMP recovery}
\label{fig:subfig}
\end{figure}

\section{Numerical Studies}
This section presents the numerical studies. To our derived thresholds
for the BOMP and ICBOMP algorithms, probabilities $p_m$ and $p_f$ are
respectively set to be 0.1\% and 0.5\%. The followings are some cited simulation results.

\subsection{on the BOMP Recovery}
The system size for the BOMP recovery is set as: $d = 50$, $N=640$ and $M=2000$.
The $N_a$ supporting blocks are chosen uniformly at random among all $N$ blocks.
Entries of the measurement matrix and the nonzero signal blocks are generated as i.i.d. complex Gaussian variables, following $ \mathcal{CN}(0, \frac{1}{M})$ and $ \mathcal{CN}(0, 1)$, respectively.
As comparisons, the thresholds in (\ref{other1}) and (\ref{other2}) are
respectively given as $\epsilon_1=0.25$ and $\epsilon_2=M \sigma^2$,
where 0.25 is a reasonable value concluded from training simulations and $M \sigma^2$ is the energy of noise vector.
The simulation results are presented as required iteration number vs.
SNR, normalized mean square error (NMSE, calculated by ${{\left\| {{\bf{s}} - {\bf{\bar s}}} \right\|_2^2} \over {\left\| {\bf{s}} \right\|_2^2}}$)
vs. SNR and successful detection probability vs. SNR,
respectively in Figure \ref{fig:subfig:a}, Figure \ref{fig:subfig:b}
and Figure \ref{fig:subfig:c}. The SNR here is
defined as $\frac{1}{\sigma^2}$. To accelerate the process, the maximum
iteration number of the BOMP is set 30 to deal with case where
the thresholds cannot stop the BOMP timely.

Figure \ref{fig:subfig:a} tells us that our derived threshold can stop the
iterations timely. As the SNR increases, the required iteration number
nearly equals to the number of supporting blocks. However, the threshold
given by (\ref{other1}) produces many false detections in low SNR regime,
and threshold given by (\ref{other2}) will make certain number of
supporting blocks missed to detect. Figure \ref{fig:subfig:b} shows that,
the NMSE achieved by the derived threshold is a little higher than that
of $\epsilon_2$ in low SNR regime, it is because some false detections
degrade the recovery performance. However it is always better than that
of set $\epsilon_1$. As the SNR increases, the output NMSE gradually becomes the smallest
among the stopping conditions. Figure \ref{fig:subfig:c} demonstrates
that, the derived threshold still guarantees a very high successful
detection probability.

\subsection{on the ICBOMP Recovery}
\begin{figure}
\centering
\subfigure[Iteration number]{
\label{fig2:subfig:a}
\includegraphics[width=2in,height=1.8in]{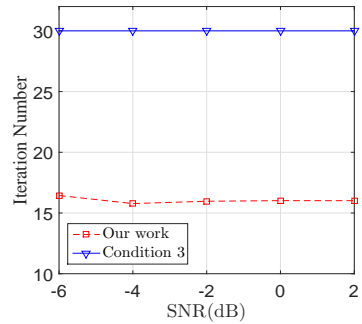}}
\subfigure[SER]{
\label{fig2:subfig:b}
\includegraphics[width=2in,height=1.8in]{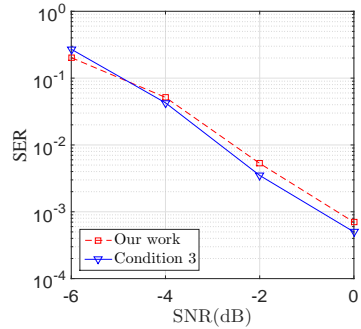}}
\caption{Performance for ICBOMP recovery}
\label{fig2:subfig}
\end{figure}

For the communication scenario in \cite{xie2015many} stated, system parameters are set:
$d=200$, $N=640$, $M_{\text{ant}}=8$, $T=5d$ and $N_a=16$, ${\rho _0}$ is the SNR.
Entries of the precoding matrices and channel vectors are generated as i.i.d. complex Gaussian variables,
following $ \mathcal{CN}(0, \frac{1}{T})$ and $ \mathcal{CN}(0, 1)$, respectively.
QPSK is applied for signal modulation. Convolutional code
is used as the error correction code and 24 bits cyclic redundancy check (CRC) code is used
as the error detection code. Soft-decision Viterbi decoding of 16 quantization levels
is used as the channel decoder. As a reference, the ICBOMP recovery will perform
30 iterations, which is exactly the case in \cite{xie2015many}. The results
required iteration number vs. SNR and symbol error rate (SER) vs. SNR
are shown in Figure \ref{fig2:subfig}.

The results show that, in the given SNR regime from -6dB to 2dB,
our derived threshold always makes the iteration number near the real sparsity level 16,
which saves nearly 14 unnecessary iterations to greatly reduce the computational cost.
In the accuracy performance, a slightly higher SER is observed for the threshold. This comes
from the fact that compared with 30 iterations, more supporting blocks will be missed to
detect when much less iterations are performed.

\section{Conclusions}
In this paper, a theoretical stopping condition is derived for greedy block sparse
recovery when the sparsity level is unknown.
By studying the energy of the residual signals at each iteration, a condition is
derived for stopping the iteration process of the BOMP algorithm.
The approach works well for the BOMP recovery. And then we promote the work
to derive the stopping condition for the ICBOMP recovery in \cite{xie2015many}.
The work contributes to saving
many unnecessary iterations.

\section*{Acknowledgment}
This work is supported by national 863 high-tech program (No.2014AA01A704),
NSF of China (No.61571412) and the Fundamental Research Funds
for the Central Universities.

\end{document}